\documentclass[conference]{IEEEtran}
\usepackage{graphicx}
\usepackage{textcomp}
\usepackage{lineno,hyperref}
\modulolinenumbers[1]
\usepackage{float}
\usepackage{siunitx}
\usepackage{booktabs}
\usepackage{multirow}
\usepackage{float}
\usepackage{color}
\usepackage{makeidx}
\usepackage{enumerate}
\usepackage{rotating}
\usepackage{epstopdf}
\usepackage{tabularx}
\usepackage{longtable}
\usepackage{amsmath}

\usepackage{enumitem}

\ifCLASSINFOpdf
\else
\fi

\usepackage{enumerate}
\usepackage{graphicx}
\usepackage{booktabs,tabularx}

\usepackage{epsfig}

\usepackage{bm}
\usepackage{mathptmx}
\usepackage{float}
\usepackage{stfloats}

\begin{document}
%
\title{A Review of Wind Speed \& Wind Power Forecasting Techniques}

\author{\IEEEauthorblockN{Harsh S. Dhiman}
\IEEEauthorblockA{Department of Electrical Engg\\
Adani Institute of Infrastructure Engineering\\
Ahmedabad, Gujarat\\
Email: harsh.dhiman@aii.ac.in}
\and
\IEEEauthorblockN{Dipankar Deb}
\IEEEauthorblockA{Department of Electrical Engg\\
	Institute of Infrastructure Technology\\Research and Management\\
	Ahmedabad, Gujarat\\
	Email: dipankardeb@iitram.ac.in}
}


%


\maketitle

\begin{abstract}
Forecasting a particular variable can depend upon temporal or spatial scale. Temporal variations that indicate variations with time, reflect the stochasticity present in the variable. Spatial variation usually are dominant in climatology and meteorology. Temporal scale for a variable can be modeled in terms of time-series. A time series is a successively ordered sequence of numerical data points, and can be taken on any variable changing with time. Wind speed forecasting applications lie majorly in the area of electricity market clearing, economic load dispatch and scheduling, and sometimes to provide ancillary support. Thus, a proper classification based on the prediction horizon i.e. the duration of prediction becomes important for various transmission system operators. 
\end{abstract}


%
\IEEEpeerreviewmaketitle

\section{Introduction}
Sustainable energy sources lead to reduction in carbon footprint and thus increase the reliability of a system \cite{bilateral,Dhiman2019a,Dhiman2020hh,SDhiman2020}. With the spurt in the installation of renewable energy sources, the fossil fuel usage is put into restricted use. Most of the industries in power sectors prefer renewable energy generation owing to its negligible carbon footprint. With wind available in abundant form, tapping power from wind is a specialized task. Considering wind as a stochastic variable, its accurate prediction can yield numerous benefits to the plant operators. Prediction involves errors, similarly in this case the error processing of forecasted wind speed/power and actual wind speed/power plays a crucial role in selecting the appropriate forecasting algorithm. \smallskip

Wind forecasting plays an important role when it comes to clearing day ahead market scenarios. Given there is a market situation to be cleared, an accurate wind forecasting scheme is helpful in such situations. Wind forecasting schemes are broadly categorized as $(i)$ Weather based prediction methods $(ii)$ Statistical or time series based prediction methods \cite{Okumus2016}. While we consider weather based prediction models,the wind forecast accuracy depends highly on the topology of the land where the wind turbines are erected. Given the topology of the land , wind speed measurements at an appropiate height from the ground, the temperature of the ambient air, air pressure etc hold important factors to take into consideration. On the other hand statistical methods solely depend on the past measurements of the wind speed to predict the future values. \smallskip

This article draws downs attention towards the recent wind energy forecasting schemes that have been in use for accurately predicting the wind power. It also lists doen several conventional statistical prediction models like persistence algorithm, ARMA and ARIMA models. Given the time scale for the forecasting, the methods can be classified based on the basis of horizon span i.e. very short-term forecasting, short-term forecasting, long-term forecasting and very-long term forecasting. Forecasting plays an important role not only in predicting accurate values of wind power but also to clear day-ahead electricity markets. It was observed that numeric weather prediction methods which are more suitable to predict wind speed for long-term durations could yield up to an average saving of 20 percent in fossil fuel\cite{Okumus2016}. Since the wind power can provide the ancillary support to the nearby power stations, forecasting accurate wind power can also help to stabilize any abnormal operation of the interconnected power plant. \smallskip

This paper provides full insight to the recent and advance wind energy forecasting schemes that can lead to optimal wind dispatch and aim to improve the economies of scale of the large power systems. The paper is organized in five sections. Section I gives an introduction to the forecasting methods, section II briefs about classification of forecasting schemes based on prediction horizon. In section III various statistical models for forecasting are discussed with persistence and ARIMA models forming the benchmark for comparing accuracy. Section IV gives insight to advance machine learning forecasting methods along with hybrid methods as well. Section V discusses various conclusions.

\section{Classification based on Prediction Horizon}
Wind speed forecasting applications lie majorly in the area of electricity market clearing, economic load dispatch and scheduling, and sometimes to provide ancillary support. Thus a proper classification based on the prediction horizon i.e. the duration of prediction becomes important for various transmission system operator's (TSO's). The time-scale classification of the wind energy forecasting methods is given as follows: \smallskip
\begin{itemize}
	\item Very short-term prediction (few seconds to 30 min
	)
	\item Short-term prediction (30 mins to 6hrs)     
    \item Medium-term prediction (6hrs to 24hrs)
    \item Long-term prediction (24hrs to 72 hrs)
    \item Very long-term prediction (72 hrs and longer)
\end{itemize}    

The above classification not only simplifies the study but also helps to choose the accurate method depending on the type of its application. Majority of the time one may find that short-term and medium term prediction methods are more in use due to its accuracy and robustness. Various performance parameters that are being used to evaluate and compare forecasting methods \cite{Okumus2016}. Given the prediction horizon the error between the forecasted value and actual value of wind speed/power can be quantified and compared using some standard error definitions such as mean absolute percentage error (MAPE), mean absolute error (MAE), mean squared error (MSE) and root mean square error (RMSE). 
The mathematical expression for the mean absolute percentage error (MAPE),mean absolute error (MAE), mean squared error (MSE) is given as follows 
\begin{eqnarray}
\label{K1}
MAPE~&=&~\frac{100}{N}\sum_{t=1}^N\frac{|p_f-p|}{p_f}\\
\label{K2}
MAE~&=&~ \frac{1}{N}\sum_{t=1}^{N}|p_f-p|\\
\label{K3}
MSE~&=&~\frac{1}{N}\sum_{t=1}^{N}(p_f-p)^2\\
\label{K4}
RMSE~&=&~\sqrt{{\frac{\sum_{t=1}^{N}(p_f-p)^2}{N}}}
\end{eqnarray}
Where in the above equations (\ref{K1}-\ref{K4}), N is the prediction horizon, $p_f$ is the is forecasted wind power and $p$ is the actual wind power. MAPE criterion for statistically analyzing the forecast accuracy has proven out to be a better parameter. The following subsections explain the various forecasting schemes (numeric weather or statistical model) that are being adopted depending on the prediction horizon, also we look the various performance evaluation parameters of the schemes.
  
\subsection{Very Short-term Forecasting Schemes}
The time-scale classification proves out to be a brownie point in terms of application of these forecasting algorithms. For any forecasting method to be accurate its performance at different sites with varied atmospheric conditions plays a crucial role in forecasting future values. Thus a parameter must be assigned that not only calculates how accurate the method is but it also determines its fitness over certain conditions. Such type of optimization studies are often done to choose the accurate forecasting method. Very short-term forecasting methods, as its name suggests predicts the future values up to a short span of time. Usual time-scale followed for this method is from a few seconds to 30 minutes. Of many methods described in literature, like spatial correlation where the 1-second ahead forecasting is done \cite{Alexiadis1999}. Artificial neural network (ANN)-Markov chain (MC) model \cite{PourmousaviKani2011} where a variable set of 175 min is taken and wind speeds for the next 7.5s ahead are predicted. Bayesian structural break model \cite{Jiang2013} are used to predict 1 min and 1 hr ahead forecasts. Data mining approach \cite{Negnevitsky2008}, . Apart from these methods, several intelligent learning methods like Artificial neural networks (ANN) in particular deep neural networks (DNN's) \cite{Hu2016} used for forecasting wind speeds up to 10 min, 30 min and 1 hr ahead. Support vector machines (SVM), hybrid methods combining Empirical Wavelet Transform (EWT) with neural networks (NN's) also form core of the very short-term wind forecasting. Recent advances in forecasting domain show evolution of machine learning algorithms, with decomposition forecasting algorithms (DFA) being most in use. The essence of a decomposition algorithm lies in breaking the time-series of a variable i.e. wind speed/power here and analyzing individual units for the forecasting and combining them to obtain the resultant series. One such decomposition algorithm is used in \cite{Zhang2016} where the time-series data of wind is broken or decomposed into several units and on each unit a feature construction process is performed and those with best features are chosen for prediction. The prediction can be carried out with standard ARIMA model of order (p,d,q) or artificial neural networks (ANN) or support vector machines (SVM). Further a forecasting technique based on Hilbert-Huang transform (HHT), that uses the decomposition technique for the non stationary and non-linear models, is also used in \cite{Liang2015}. 

\subsection{Short-term Forecasting Schemes}
Short-term wind forecasting is the most used forecasting category with many day-ahead markets need to clear the market scenarios by the end of the day. Among these the most used forecasting methods are a combination of two or more machine learning methods combined with a time-series model (AR, MA, ARMA,ARIMA,ARMAX). ARMAX stands for autoregressive exogenous moving average, is a non-linear model that captures all the uncertainties that are related to the stochastic nature of the wind. Among various statistical models \textit{ARMA} model is a popular forecasting method. In \textit{ARMA}, AR stands for auto-regressive and MA stands for moving average. Thus a a combined autoregressive (of order \textit{p}) and moving average (of order \textit{q}) forms a ARMA (\textit{p,q}) model. The order of ARMA models i.e. \textit{p} for AR and \textit{q} for MA denote the lag between present and past values of the variable under test. Thus a generic \textit{ARMA} model can be mathematically expressed as follows: 
\begin{eqnarray}
	\label{IM2}
	(y_t-\mu_x)&=&\alpha_1(y_{t-1}-\mu_x)+\alpha_2(y_{t-2}-\mu_x)+.....+\nonumber\\
	&&\alpha_p(y_{t-p}-\mu_x)+\xi_t-\beta_1\xi_{t-1}-\beta_2\xi_{t-2}-...\nonumber\\
	&&-\beta_q\xi_{t-q}
\end{eqnarray} 
where $\xi_t$ is an independent process with mean zero and $\alpha_1,\alpha_2,...,\alpha_p$ and $\beta_1,\beta_2,...,\beta_q$ are the parameters of the AR and MA process respectively. \smallskip 

In \cite{Kavasseri2009}, a variant of ARIMA model, i.e. \textit{f}-ARIMA is used to predict the day ahead wind forecasts. \textit{f} stands for fractional, where the value of differencing parameter $d\in(-0.5,0.5)$. Results of \cite{Kavasseri2009} show that \textit{f}-ARIMA model was 42\% more efficient than persistence model. Apart from these standard ARIMA models, a hybrid model i.e. wind forecasting by wavelet transform and neural networks used in \cite{Catalo2011}, to predict the wind power for 3 hr ahead. Data from previous 12 hrs with 15 min time step was taken as input to the ANN's input layer. The original wind power series is decomposed using wavelet transform (specifically D-WT), and the resulting series is fed to the neural network where the future values are forecasted. Wavelet transform also find their use in forecasting the load and electricity prices for a power plant particularly in deregulation market \cite{Conejo2005}.\smallskip

Another short-term forecasting method that involves empirical mode decomposition (EMD) and feature selection was studied in \cite{Zhang2016}, where the wind series was broken down into several subsequent series. Each of these series, an intrinsic mode function (IMF) was computed. The IMF of a decomposed signal represents the irregularity and frequency components of the signal. Once the series is decomposed the appropriate forecasting tool i.e. ANN or SVM is chosen to forecast the wind power values. A generic expression of EMD based decomposition can be represented as follows:
\begin{equation}
\label{K5}
{y_t}=\sum_{i=1}^{N}{c_i}(t)+{r_n(t)}
\end{equation}

where in the above equation \ref{K5} $c_i(t) (i=1,2,...,N)$ are the different IMF's and $r_n(t)$ is the final sum of all the residuals.\smallskip 

In \cite{Mohandes2004}, support vector machine technique for wind prediction is used. The idea behind carrying out such procedure is to map out time-series data of any variable into higher dimension space (for e.g. hilbert space) and carrying out regression analysis. Mathematically SVM can be expressed with the following set of equations:
\begin{eqnarray}
\label{K6}
f(x)~&=&~\sum_{i=1}^{D}w_i\phi_i(x)+b\\
\label{K7}
R[w]~&=&~\frac{1}{N}\sum_{i=1}^{N}|f(x_i)-y_i|_\epsilon+\lambda\|\|w\|\|^2
\end{eqnarray}

In the above equations (\ref{K6}-\ref{K7}), $\phi_i(x)$ are features and $w_i$ and $b$ are the coefficients to be computed from the data sets. The coefficients $w_i$ can be found out by minimizing the function $R[w]$. Also the results of the prediction obtained from SVM were compared with multi-layered perceptron (MLP). The performance measures used in \cite{Mohandes2004} were MSE and RMSE. It was found that SVM performed better than MLP as former had a mean squared error of 0.78\%  compared to 0.9\% of latter. Many studies have also been carried out that compare the performance of MLP with standard AR process of order \textit{p}. Another method where a combination of wavelet transform, support vector machines and genetic algorithm was used in \cite{Liu2014}. As mentioned in previous methods \cite{Zhang2016}, a wind series is decomposed using wavelet transform. The process of genetic algorithm helps to select the input parameters for the SVM. It is necessary to have optimized input in order to choose the best candidates for the forecast. The results showed that the MAPE obtained with WT-SVM-GA was around 14.79\% and that with persistence was 22.64\%. Following figure shows the results of forecasting with WT-SVM-GA method \cite{Liu2014}. In the fig \ref{IM1} below it can be seen that the technique used tracks the actual data sets accurately and precisely.  
\smallskip

\begin{center}
	\vspace{-0.4cm}
	\begin{figure}[h]
		\centering
		\includegraphics[width=0.92\linewidth]{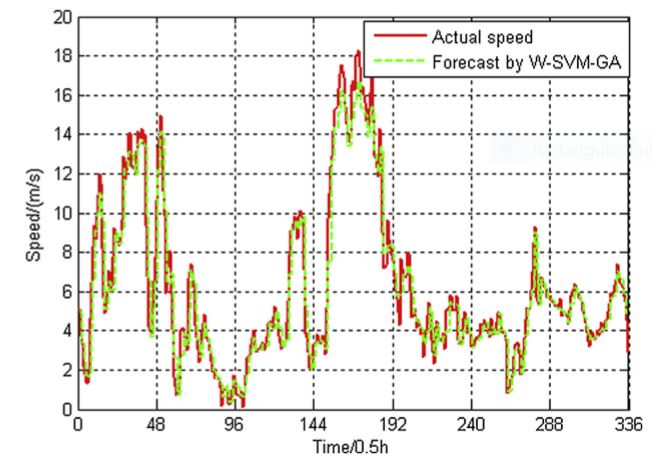}
		\vspace{-0.3cm}
		\caption{Wind speed forecasts with WT-SVM-GA.}
		\label{IM1}
		\end{figure}
		\vspace{-0.1cm}
	\end{center}

\subsection{Long-term \& Very Long-term Forecasting Schemes}
Long-term forecasts find their major application in the field of unit commitment decisions, maintenance scheduling etc. In \cite{Pousinho2011}, a hybrid method combing particle swarm optimization (PSO) and adaptive-network-based fuzzy interference system (ANFIS) i.e. PSO+ANFIS was used for forecasting 1-day ahead in the time steps of 15 min. The PSO algorithm was chosen to find out the best parameters for neuro-fuzzy systems. Another method \cite{Fei2015}, where wavelet decomposition (WD) and artificial bee colony optimization (ABCO) based relevance vector machine (RVM) is used. Here a wind signal is decomposed into different sub-series of different frequency ranges and then further the forecasting of different models is done by RVM. The kernel parameters of RVM were chosen by a meta-heuristic algorithm known as artificial bee colony optimization (ABCO). The forecast model for relevance vector machines can be mathematically expressed as follows:
\begin{eqnarray}
\label{K8}
h_n~&=&~y(x_n,w)+\varepsilon_n\\
\label{K9}
y(\textbf{x,w})~&=&~\sum_{i=1}^{N}w_iK(x,x_i)+w_0
\end{eqnarray}     

In the above equations (\ref{K8}-\ref{K9}), $x_n$ is the input vector of the decomposed wind series, $\varepsilon_n$ is added noise in the process, $K(x,x_i)$ is the kernel function which can be expressed as follows:


Apart from the above mentioned forecast techniques, many numeric weather methods such as Global forecast system \cite{SalcedoSanz2009}, fifth generation mesoscale model (MM5) with neural network was used. The predictions from the global forecast systems (GFS), plus the atmospheric conditions of the topography concerned are used as boundary conditions for the MM5 model. Its output is given as an input to the neural network. From the global forecast method, the physical downscaling of the wind data is done, then using the MM5 model, the statistical downscaling is achieved.  

\section{Statistical Models for Wind Forecasting}
Wind speed/power forecasting plays an important role for a TSO in order to guarantee certain amount of power transfer to the grid. In literature among various prediction models, time-series model \cite{CROONENBROECK2015201} outperform numeric weather prediction model and its general mathematical expression could be written as $y_t=y_{t-1}+\epsilon$ i.e. the future values of a variable $y$, here wind speed/power in our case, depend on its past values $y_{t-1}$ \cite{Okumus2016}. The most simple and effective method for wind speed/power prediction is persistence method where the data from the past is used to predict the variable under test. It also important to note that wind power forecasting yields better forecast results than wind speed forecasting. The wind speed is mapped into wind power by following the power-speed curve for a particular wind turbine. For any new forecasting algorithm developed, persistence method acts as benchmark for accuracy determination. 

\subsection{Regressive Models for Wind Energy Forecasting}
Statistical models have proved out to be more efficient and useful when it comes to short-term and medium term forecasting. A lot of data set-points are to be needed to forecast the future values of wind speed/power. The most commonly used statistical approach is AR, MA, ARMA \& ARIMA models. These models have performed decently for short-term and medium-term forecasts. ARIMA models have been in use while considering hybrid or combinational methods for short-term forecasts. In \cite{Lydia2016}, linear and non-linear regressive models have been discussed for short-term forecasting. The general expression for linear auto-regressive moving average with exogenous input (ARMAX) model can be represented by the following equation:
\begin{eqnarray}
\label{K10}
A(q)y(t)~&=&~B(q)u(t)+C(q)e(t)\\
\label{K11}
A(q)~&=&~1+a_1q^{-1}+a_2q^{-2}+...+a_pq^{-p}\\
\label{K12}
B(q)~&=&~1+b_1q^{-1}+b_2q^{-2}+...+b_fq^{-f}\\
\label{K13}
C(q)~&=&~1+c_1q^{-1}+c_2q^{-2}+...+c_lq^{-l}
\end{eqnarray} 
In the above equations (\ref{K10}-\ref{K13}), $A(q)$, $B(q)$, $C(q)$ are the AR, exogenous and MA parts respectively. Whereas $p$, $f$ and $l$ are orders of the AR, exogenous and MA parts. Whereas non-linear regressive models \cite{Lydia2016} can be expressed as follows:
\begin{eqnarray*}
\label{K14}
y(t)=f(y(t-1),y(t-2),..y(t-p), u(t-1),u(t-2)\\
,...u(t-f), e(t-1), e(t-2),..e(t-l))
\end{eqnarray*}   

In the above equation \ref{K14}, $y(t)$ are the future values, and $e$ represents white noise component. ARMAX models involve exogenous variable, its origin is from the meteorological changes in the wind movement \cite{book1}. The non-linear ARMAX models have been derived from various machine learning algorithms i.e. support vector machine (SVM), M5R and bagging. Support vector machines transform the data-sets from one dimension to higher dimension. Let $x$ be the input space vector then, $z=\phi(x)$ is the feature space vector and $\phi$ is the function that maps $x$ to $z$.\smallskip

In  \cite{Lydia2016}, linear regression models were built in MATLAB using 10-min data sets from past 15 days. The results showed that the ARMAX models with wind direction as exogenous input (for e.g wind speed at different heights, wind direction, temperature and solar radiation) performed better than rest of the models. The performance criterion used here were MAE, RMSE and MAPE. The figure \ref{IM2} below shows the wind speed prediction using ARMAX$(2,3,2)$ non-linear regression model based on SVM, where $p=2$, $f=3$ and $l=2$ are the orders of AR, exogenous input and MA part respectively. Two datasets have been taken, one from Sotaventa Galicia Plc a wind farm experiment supported by regional autonomous government and dataset 2 from M.S Puram, Madurai which is one of the monitoring station of Centre of Wind Energy Technology (CWET), Chennai.    

\begin{center}
\vspace{-0.4cm}
\begin{figure}[h]
	\centering
	\includegraphics[width=0.92\linewidth]{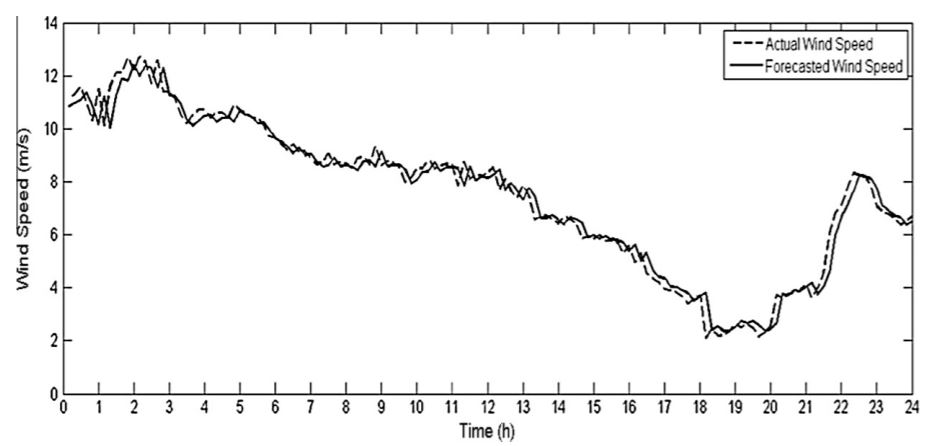}
	\vspace{-0.3cm}
	\caption{10-min ahead wind speed forecast using ARMAX(2,3,2) \cite{Lydia2016}.}
	\label{IM2}
\end{figure}
\vspace{-0.1cm}
\end{center} 

\subsection{ARMAX based Forecasting Scheme}
Autoregressive moving average exogenous input (ARMAX) based linear regressive models are often tested for wind speed/power forecasting. One such forecasting was done for Ashcroft, one of the villages in British Columbia, Canada. The historic hourly data set of first 15 days of October'17 was taken and ARMAX model was formulated for the same. The ARMAX model chosen was ARMAX(2,2,2) i.e. with reference to equation (\ref{K11}-\ref{K13}) the orders of AR, MA and exogenous part were 2,2,2 respectively. The equation below describes the discrete-time ARMAX model for the same:
\begin{eqnarray}
A(z)y(t)~&=&~B(z)u(t)+C(z)e(t)\\
A(z)~&=&~1-\frac{0.3659}{z}+\frac{0.09601}{z^2}\\
B(z)~&=&~\frac{0.3659}{z}+\frac{0.09601}{z^2}\\
C(z)~&=&~1-\frac{0.03659}{z}+\frac{0.03175}{z^2}
\end{eqnarray}

In the above equations, $y(t)$, $u(t)$, and $e(t)$ represent the output variable wind speed, exogenous variable and added white noise respectively. The MSE using ARMAX(2,2,2) was found to be 8.805\%. 

\begin{center}
	\vspace{-0.4cm}
	\begin{figure}[h]
		\centering
		\includegraphics[width=0.92\linewidth]{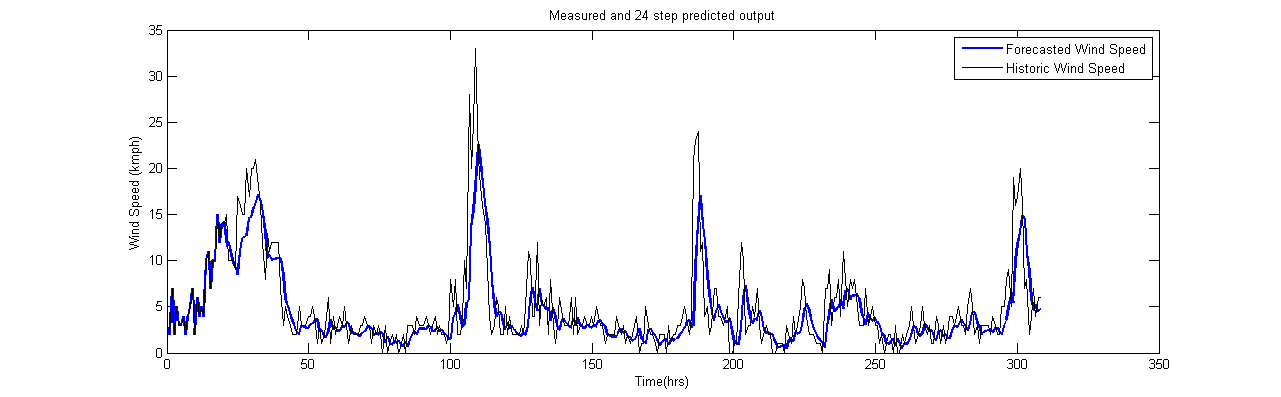}
		\vspace{-0.3cm}
		\caption{Wind speed forecast using ARMAX(2,2,2) for Ashcroft, BC, Canada}
		\label{IM8}
	\end{figure}
	\vspace{-0.1cm}
\end{center}

\section{Machine Learning based Forecasting Algorithms}
This section gives an insight to the various machine learning based forecasting schemes that are being employed globally for accurate wind forecasts. Machine learning methods like Artificial Neural Networks (ANN'S,), Support vector machine (SVM), Empirical mode decomposition (EMD), Support vector regression (SVR) and Extended machine learning (ELM). Ensemble methods \cite{REN201582} . The subsections below describe each of the above mentioned methods in detail.

\subsection{Forecasting with Artificial Neural Networks (ANN's)}
Wind speed forecasting helps in utilizing the available wind resource to its optimum capacity. This not only reduces the carbon footprint in the ecosystem but also facilitates the possibility of interconnection of power systems. In terms of forecasting, various machine learning algorithm have come up that attempt to solve the problem of non linearity in prediction models. Artificial Neural Networks or just neural networks are one such technique. Given any data set, ANN's learn from the experience of the past and predict future values. ANN's find their application not only in wind forecasting but also in wind turbine control \cite{Ata2015}. The basic working of an artificial neural comes from the working of brain i.e. communication via neurons \cite{book2}. The figure \ref{IM3} shows the multi-layer architecture for an ANN. It contains an input layer, hidden layer(s) and an output layer. The input layer of the ANN has a weight associated with it. Mathematically ANN's can be expressed as follows:

\begin{center}
	\vspace{-0.4cm}
	\begin{figure}[h]
		\centering
		\includegraphics[width=0.92\linewidth]{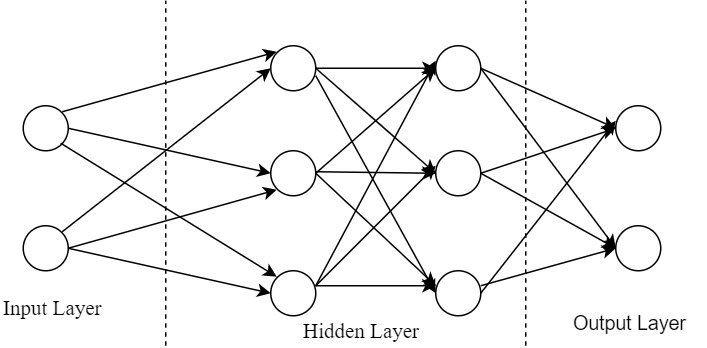}
		\vspace{-0.3cm}
		\caption{Multi-layer ANN architecture}  
		\label{IM3}
	\end{figure}
	\vspace{-0.2cm}
\end{center}

\begin{eqnarray}
\label{K15}
f(x)~&=&~\phi(\sum_{i=1}^{N}w_ix_i+b)\\
\label{K16}
f(x)~&=&~\phi(\textbf w^Tx+b)\\
\label{K17}
\phi(g)~&=&~\frac{1}{1+e^{-g}}
\end{eqnarray}

In the above equations (\ref{K15}-\ref{K16}), $w_i's$ are the weights associated with each neuron present in the input layer, $x_i's$ are the inputs and $b$ is the bias term. $\phi$ in equation \ref{K17} denotes the activation or transfer function (here sigmoid function). The output from the weighted sum of the neurons is then sent to the activation function block where a mathematical function is used to scale down the output to zero or one. Several combinations of ANN architecture's are possible by keeping different number of hidden layers, different number of neurons in each layer and the choice of activation function. ANN's find their application mostly in short-term wind energy forecasting \cite{Ata2015}. The following are the hybrid systems of ANN's 

\begin{itemize}
	\item Genetic Algorithm (GA) \& Neural Networks
	\item Particle Swarm Optimization (PSO) \& Neural Networks     
	\item Wavelet Neural Networks (WNN's)
	\item Fuzzy Neural Networks (FNN's)
\end{itemize} 

The following sections describe the application part of ANN's to various categories of wind speed/power forecasting as mentioned in section II i.e. based on prediction horizon. \smallskip    
       
\subsubsection{ANN's in Short-term wind forecasting}
Artificial neural networks are heavily used to forecast wind speed/power for a duration ranging from 30-mins to 6 hrs. Among ANN's the most commonly used topology is multi-layered perceptron (MLP). In \cite{Catalo2011}, short-term wind power forecasting for wind farms in Portugal is achieved via a hybrid method of wavelet-transform and neural networks (NN). Here first the wind speed data is broken down into sub-series via discrete-wavelet transform and later is fed to NN for the training part. As mentioned earlier the number of neurons in each layer can affect the performance, the neurons in each layer can be chosen by trail or error method. Once the training algorithm is through, the data sets are sent for learning stage where the error minimization takes place between input values and desired values. Usually backpropagation is used as learning algorithm \cite{7974446}. However backpropogation is a slower technique, it is replaced by Levenberg-Marquardt algorithm. Catalao \textit{et al} \cite{Catalo2011} have compared the NNWT approach with ARIMA (1,2,1) and NN where the forecast for 3-hr ahead is done by taking historical data of previous 12-hr. The MAPE value using NNWT approach was found to be 6.97\%. \smallskip

\subsubsection{Wind Forecasting using Recurrent Neural Networks} \smallskip
In \cite{Marhon2013} talks about the architecture of a recurrent neural network. Most of the neural network based forecasting schemes employ MLP architecture to forecast wind power. However in RNN, there are connections of processing elements (PE) i.e. neurons from output layer to preceding layer (hidden layer), or subsequent layers and every connection has a weight associated with it. MLP is a special case of RNN where the weight coefficients are zero. The figure \ref{IM4} below gives a diagrammatic representation of recurrent neural networks. Mathematically RNN's can be described as follows:
\begin{center}
	\vspace{-0.2cm}
	\begin{figure}[h]
		\centering
		\includegraphics[width=0.92\linewidth]{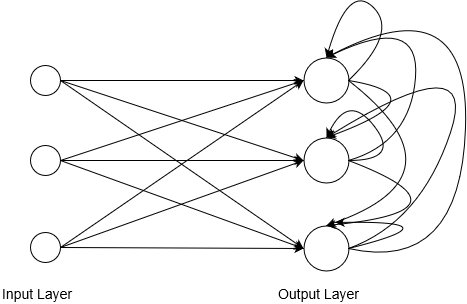}
		\vspace{-0.3cm}
		\caption{RNN architecture}  
		\label{IM4}
	\end{figure}
	\vspace{-0.2cm}
\end{center}

In \cite{CAO2012148} a comparative analysis of univariate and multivariate ARIMA models is done with RNN's. The wind speed data was acquired from Wind Engineering Research Field Laboratory (WERFL) at Texas Tech University. The wind speed was recorded at 5 different heights (8,13,33,70,160 ft) can the forecasting was done. It was found that the forecast accuracy improved at higher heights from ground. Various models like univariate ARIMA, univariate RNN, multivariate ARIMA and multivariate RNN were examined and it was found that RNN's outperform ARIMA models by and large. 

\subsection{Support Vector Machines}

Support vector regression (SVR) is a type of Machine learning regression which is associated with learning algorithm equipped to analyze historical data for classification and regression. SVR works on the principle of structural risk minimization (SRM) from statistical learning theory \cite{VapLer63}, \cite{Cortes1995}. The core idea of this theory is to construct a hypotheses $\textit{h}$ that yields lowest true error for the unseen and random sample testing data \cite{Joachims1998}. Apart from SVR, a universal machine intelligent technique like Artificial neural network (ANN) with applications in character recognition, image compression and stock market prediction, is studied \cite{Schlkopf1997}. Shirzad et al. have compared the performance of ANN and SVR to predict the Pipe Burst Rate (PBR) in Water Distribution Networks (WDNs) \cite{Shirzad2014}, and found ANN to be a better predictor than SVR, but generalization is not consistent with physical behavior. SVR has an advantage over ANN with respect to the number of parameters involved in training phase. The computation time is another important factor for carrying out regression analysis.

Consider a set of training data (historical data) ${(x_1,y_1),\dots,(x_n,y_n)}$ $\subset X \times \mathrm{R}$, where $X$ denotes the input feature space of dimension $\mathrm{R^n}$. Let $Y= (y_1,y_2,\dots,y_i)$ denote the set representing the training output or response, where $i=1,2,\dots,n$ and $y_i \in \mathrm{R}$.      

\subsubsection*{$\varepsilon$-support vector regression}
This type of SVR uses an $\varepsilon$-insensitive loss function that intuitively accounts for sparsity similar to SVRby ignoring errors less than $\varepsilon$. 

$\varepsilon$-SVR aims to find a linear regressor \begin{align}
f(x)=w^{T}x+b, with \hspace{0.1cm} w\in X , b\in \mathrm{R},
\end{align}
for prediction, where $x \in X$ is the input set containing all the features, $w$ is the weight coefficient related to each input $x_i$ and $b$ is the bias term.

The objective is to find the $f(x)$ with maximum deviation $\varepsilon$ from the respective feature sets while being as flat as possible. In order to achieve the flatness of the desired regressor, the square of the norm of weight vector $w$ needs to be minimized, and the SVR problem is structured in the form of a convex optimization problem \cite{Vapnik2000} given as
\begin{align}
\text{min} \hspace{0.1cm} \frac{1}{2} \parallel w\parallel^2 + C(e^T\chi+e^T\chi^*),\\
\textbf{subject to} \hspace{0.2cm} y-w^{T}x-eb \le e\varepsilon +\chi, \chi \ge 0,\\ \nonumber
w^{T}x+eb-y \le e\varepsilon +\chi^{*}, \chi^* \ge 0,
\end{align}     
where $C$ is the regularization factor that reflects the trade-off between the flatness of regressor $f(x)$ and the maximum tolerable deviation $\varepsilon$. The value of $\varepsilon$ introduces a margin of tolerance where no penalty is imposed on the errors. The larger $\varepsilon$ is, the larger are the errors. The parameter $C$ controls the amount of influence of the error. The variables $\chi, \chi^*$ are the slack variables introduced as a soft margin to the tolerable error $\varepsilon$ and $e$ is a vector of ones of appropriate dimensions ($n\times1$).

\begin{figure}[h]
	\centering
	\includegraphics[width=1\linewidth]{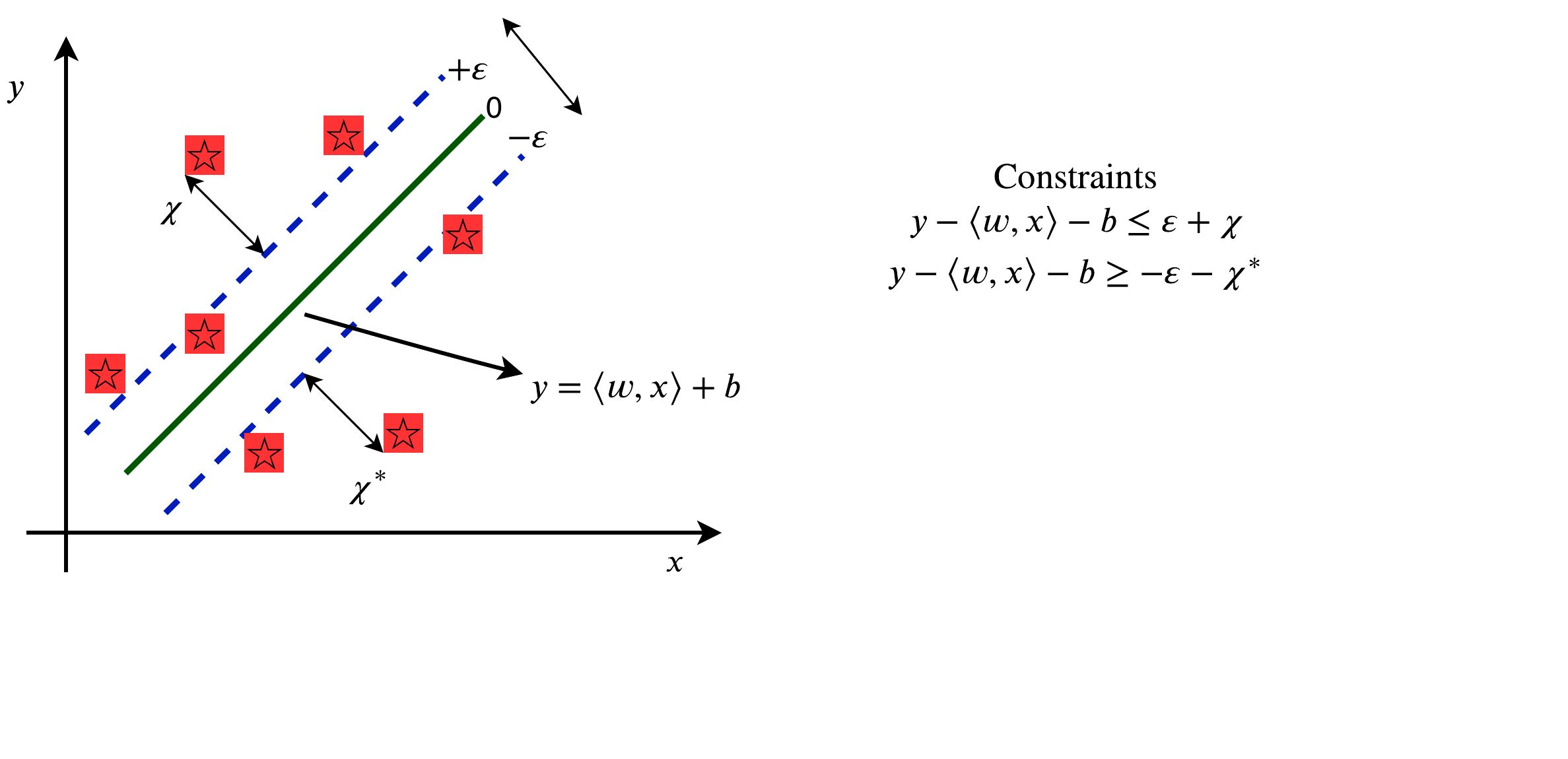}
	\vspace{-1.8cm}
	\caption{Diagrammatic explanation of $\varepsilon$-SVR }
	\label{fig:svrexp}
	\end{figure}

In machine learning regression, the problem of over-fitting persists which results in less error in training phase and high error in testing phase. Commonly used regularization techniques include $\mathcal{L}_1$ and $\mathcal{L}_2$ regularization. Mathematically, these are expressed as
\begin{align}
\mathcal{L}_1 &\colon \arg \min_w ~ \text{loss function} + \lambda\sum_{i=1}^{n}|w_i|\\
\mathcal{L}_2 &\colon \arg \min_w~ \text{loss function} + \lambda\sum_{i=1}^{n}w_i^2 
\end{align}
The formulation of SVR is diagrammatically explained in Figure \ref{fig:svrexp} where the ``stars" represent the support vectors, green solid line shows the SVR regressor and blue dashed lines are the hyperplanes with soft limit on the tolerance error $\varepsilon$.  

	However, this is not the case always, as the feature sets might not be linearly separable. To handle such nonlinearities in the feature sets, kernel trick or often called as kernel functions are used to transform data to a higher dimensional space. After transformation via suitable mapping function $\phi: \mathrm{R^n}\rightarrow Z$, the data becomes linearly separable in the target space (high dimensional space), that is, $Z$. The inner product $\langle w^{T},\phi(x)\rangle$ in the target space can be represented by using kernel function. Kernel functions are similarity functions which satisfy Mercer's theorem such that $k(x_i,x_j)= \langle\phi(x_i),\phi(x_j)\rangle$, are the elements of the kernel matrix $K$. Several kernel functions are available in literature like linear, polynomial with degree $d$, Gaussian, Radial Basis Function (RBF) with bandwidth of the function $\sigma$ and exponential function.
	
\begin{figure}[H]
		\centering
		\includegraphics[width=1.0\linewidth]{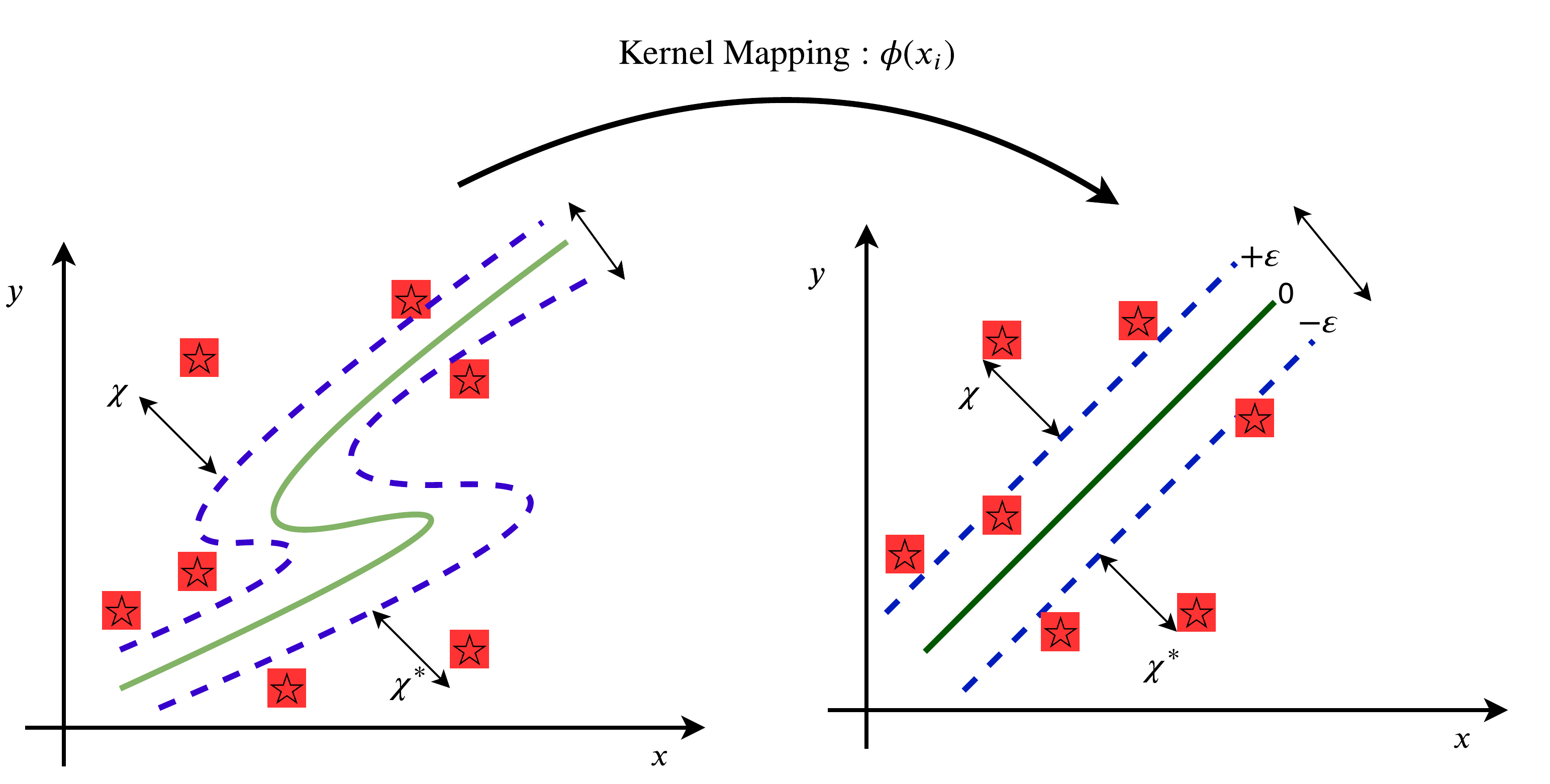}
		\caption{Kernel mapping into higher dimension space for non-linear datasets}
		\label{fig:kerneltrick}
		\end{figure}

\begin{figure}[H]
			\centering
			\includegraphics[width=0.81\linewidth]{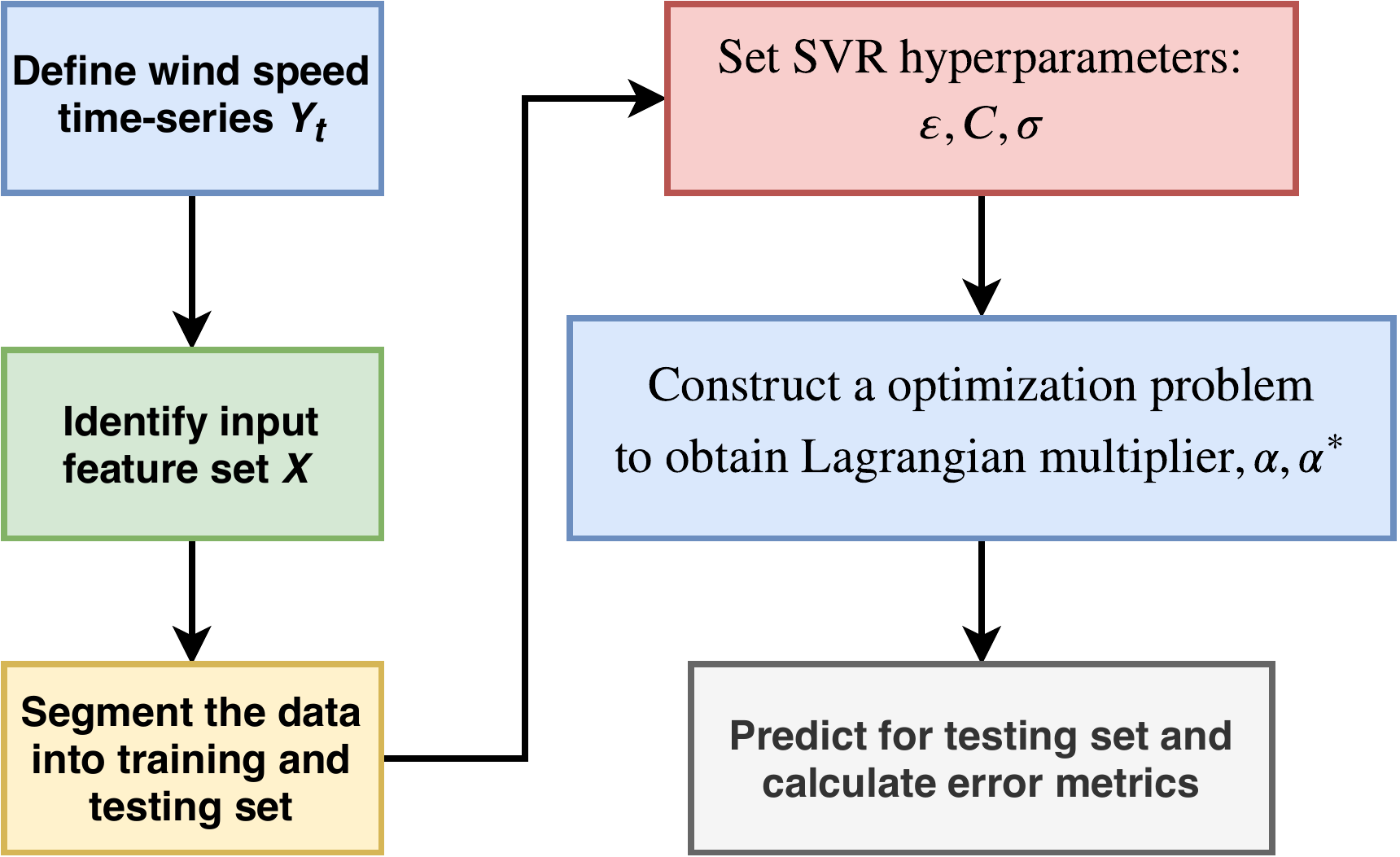}\vspace{0.1cm}
			\caption{Schematic flowchart for $\varepsilon$-SVR algorithm}
			\label{fig:svrflowchart}\vspace{-0.05cm}
			\end{figure}

	Figure \ref{fig:kerneltrick} illustrates the kernel trick used when the input vectors are not linearly separable. This transformation makes the computation of weights and bias vector much easier. 
		Now, we look at the dual form of the optimization problem as
		\begin{align}
		\text{min} \hspace{0.1cm} \frac{1}{2}\sum_{i,j=1}^{n}(\alpha_i-\alpha_i^{*})^{T}k(x_i,x_j)(\alpha_j-\alpha_j^*)+e^T\varepsilon\sum_{i=1}^{n}(\alpha_i+\alpha_i^*)\nonumber\\-\sum_{i=1}^{n}y_i(\alpha_i-\alpha_i^*) \\
		\textbf{s.t.} \hspace{0.3cm} e^T\sum_{i=1}^{n}(\alpha_i-\alpha_i^*) = 0,~~ 0\le\alpha,~~ \alpha^*\le Ce, \nonumber
		\end{align}
		where $\alpha$ and $\alpha^*$ represent positive and negative Lagrange multipliers such that $\alpha_i\alpha_i^*=0$,~ $i=1,2,\dots,n$. The regressor $f(x)$ can be written as
		\begin{align}
		\label{e5}
		f(x)=\sum_{i=1}^{n}(\alpha_i-\alpha_i^*)k(x,x_i)+b.
		\end{align}
		The complexity of this regressor is independent of the dimensionality of the feature set but only depends on the number of support vectors which are nothing but the data points which separate the feature sets from each other. However, the performance of the SVR also depends on the choice of kernel function and helps in reducing the computation time of the regression. A flowchart with a step-by-step implementation of the $\varepsilon$-SVR algorithm is described in Fig. \ref{fig:svrflowchart}.
		
\smallskip
\section{Hybrid Algorithms for Wind Forecasting}  
We have seen the various forecasting schemes each with their own set of advantages and being superior to others in terms of performance metrics i.e. RMSE, MSE and MAPE. However, the focus has been shifted recently on the hybrid forecasting schemes which combine two or more prediction method and are aggregated thereafter. Hybrid methods like ANN-ARIMA \cite{CADENAS20102732} , PSO-ANFIS \cite{POUSINHO2011397}, Wavelet transform-Neural Network (WT-NN) \cite{Catalo2011}, Kalman filter-ANN (KF+ANN) \cite{SHUKUR2015637} , Wavelet-Support vector machine optimized by genetic algorithm (WT-SVM-GA) \cite{LIU2014592} , Empirical mode decomposition-support vector machine (EMD+SVM) \cite{Zhang2016}, Fast Ensemble Empirical mode decomposition- regularized extended machine learning (FEEMD-RELM) \cite{SUN2016197} etc. 

Hybrid forecasting methods have been in trend recently owing to its additional advantages over conventional single forecasting methods. Wind being a stochastic variable, its non-stationary and non-linear characteristics cause  difficulties in its prediction \cite{BABU201427}. Also statistical models have outperformed numerical weather predictions models like  mseo-scale modeling with MM5 and Global Forecast System \cite{JUNG2014762}. With the growing energy demand, operators these days need to ensure that their forecast errors are minimum, which not only calls for the need of accurate methods but also captures the non-linearity oif the wind speed. Decomposition algorithms like empirical mode decomposition and wavelet transform \cite{Catalo2011} in combination with time-series models like ARIMA and ARMAX have been in use recently. Machine learning methods like ANN and SVM have also found to be beneficial to address the non-linearity. Empirical mode decomposition involves decomposing the original wind speed data sets into subsequent subsets, performing forecasting on each of the subsets and further aggregating the results for final forecast. Individual forecasts could be based on time-series prediction models or machine learning algorithms. In \cite{Zhang2016} the author uses decomposition algorithm on original wind speed series. Further each sub series is forecasted using ANN or SVM. The figure belows gives a diagrammatic representation of forecasting process using EMD and feature selection performed on a wind speed or power time series.
\begin{center}
	\vspace{-0.2cm}
	\begin{figure}[h]
		\centering
		\includegraphics[width=0.92\linewidth]{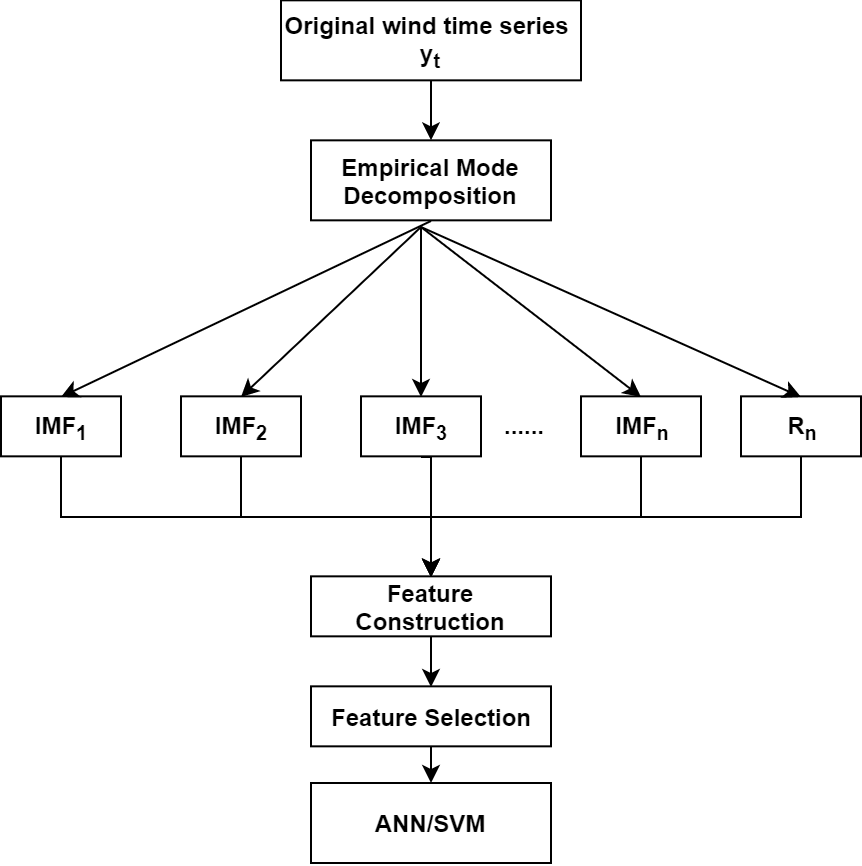}
		\vspace{-0.3cm}
		\caption{EMD based wind speed forecasting methodology \cite{Zhang2016}}  
		\label{IM7}
	\end{figure}
	\vspace{-0.2cm}
\end{center}
The framework for the given decomposition process can be stated as follows:
\begin{enumerate}
	\item The wind speed series is decomposed by Empirical Mode Decomposition to get IMF's and a residual.
	\item The feature construction takes place that collects information from all the IMF's and the residual.
	\item The optimal feature subset is selected from feature combinations.
	\item The forecasting model is built using ANN or SVM from the feature subsets. 
\end{enumerate}
EMD and feature selection technique \cite{Zhang2016} is used to select the best features from the resulting sub series. Given a wind speed series it can be expressed after decomposition as follows:
\begin{eqnarray}
\label{K32}
\textbf{x}&=&[c_1(t),c_2(t),...,c_n(t),r_n(t),c_1(t-1),c_2(t-1)...\nonumber\\&&c_n(t-1),r_n(t-1),c_1(t-p+1),c_2(t-p+1),...\nonumber\\&&c_n(t-p+1),r_n(t-p+1)]
\end{eqnarray} 
In the above equation the original time series $y_t$ is decomposed using EMD and the linear regression of order $p$ is applied to obtain the feature vector \textbf{x}. The study is done on three wind farms in china (Jiangsu, Ningxia and Yunnan). Data sets for one month during are taken to evaluate the forecasting model. The following table shows the error metrics for the three wind farms.
\begin{table}[h]
	\caption{Performance metrics for Jiangsu, Ningxia \& Yunnan using DSF-ANN algorithm \cite{Zhang2016}}
	\label{tab:Gradient}
	\begin{center}   
		\begin{tabular}{|l|l|l|}
			\hline
			Wind Farm & RMSE (m/sec) & MAPE(\%) \\\hline
			Jiangsu & $0.80$ & $17.98$ \\\hline
			Ningxia & $0.80$ & $10.29$ \\\hline
			Yunnan & $0.82$ & $11.25$ \\\hline
		\end{tabular}
	\end{center}
\end{table} 

In \cite{6979269}, wind forecasting is done with an improved version of EMD i.e. ensemble EMD (EEMD), complementary ensemble EMD (CEEMD) and complete ensemble EMD with adaptive noise (CEEMDAN). However authors report that though EMD appeared to outperform various conventional methods (Persistance \& ARMA), there appeared to be a problem of mode mixing \cite{norden} i.e. multiple IMF's contain signals in similar frequency band. EEMD algorithm is similar to EMD except that the input time series has additional finite gaussian white noise. Othe versions of EMD like CEEMD and CEEMDAN proved out to be better than its parent version EMD. Results of \cite{6979269} show that EMD based hybrid SVR methods outperformed EMD based hybrid ANN methods for 1-hr, 3-hr and 5-hr ahead forecasting.Another hybrid method \cite{Catalo2011} is tested for short-term wind forecasting where wavelet transform and neural networks are used for 3-hr ahead prediction. Wavelet transform converts a wind series into several sub series, each sub series has a better performance behavior than the original one owing to the filtering effect of wavelet transform. Wavelet transform has two categories i.e. continuous wavelet transform and discrete wavelet transform (DWT). DWT can be mathematically defined as: 
\begin{eqnarray}
\label{K23}
W(m,n)=2^{-(m/2)}\sum_{t=0}^{T-1}f(t)\phi\left(\frac{t-n.2^m}{2^m}\right)
\end{eqnarray}    
In the above equation \ref{K23}, $T$ represents the length of the signal $f(t)$ , where $m$ and $n$ are the integer variables, and $t$ is the discrete time index \cite{1388509}.

Research in the field of wind forecasting has increased tremendously for the hybrid models based on machine learning \cite{Dhiman2019}. Candenas and Rivera demonstrated ARIMA-ANN based forecasting model where for a fixed prediction horizon, wind forecasting is carried out \cite{Cadenas2010}. Liu et al. presented a Support vector machine and Genetic algorithm (GA) based hybrid forecasting method using Wavelet decomposition transform for fragmenting the wind speed time-series to eliminate any potential stochastic variation \cite{Liu2014}. Zhang et al. presented a hybrid technique based on gaussian process regression (GPR) and  auto-regression (AR) and compared their results with SVM, ANN and persistence algorithm \cite{Zhang2016}. Mi et al. explored a hybrid method that encompasses wavelet decomposition transform, extreme learning machine and outlier correction technique to forecast multi-step wind speed \cite{Mi2017}. Wavelet and wavelet packet decomposition eliminates noisy component from the wind series and extreme learning machine provides multi-step forecast on the sub-signals obtained from the decomposition technique.

Li et al. presented a combined method based on constant weight and variable weight for short-term wind speed prediction  \cite{Li2018}. Jiang et al. proposed a novel method for short-term wind prediction based on modeling the fluctuations caused by adjacent wind turbines and the selected inputs are given to a v-SVM model \cite{Jiang2017}. Azimi et al. presented a feature selection model based on k-means cluster and a multilayer perceptron neural network for predicting short-term wind speed \cite{Azimi2016}. Jiang et al. presented correlation based discrete wavelet transform (DWT), least-square support vector machine (LSSVM) and generalized autoregressive conditional heteroscedastic (GARCH) method for short-term wind speed prediction. Correlation coefficients among different sub-series are used to assess the inputs to the LSSVR models for wind speed prediction  \cite{Jiang2018}. 

Liu et al. discussed a modified  Broyden-Fletcher-Goldfarb-Shanno (BFGS) neural network and wavelet transform based signal processing technique for short-term wind speed prediction and validated the same for four wind speed datasets \cite{Liu2018}. Correlation coefficients are determined for each sub-series obtained after wavelet decomposition for assessing their relative importance. Tian et al. presented a multi-objective forecasting algorithm \cite{Tian2018} wherein data pre-processing technique is based on complementary ensemble empirical mode decomposition (EEMD), variational mode decomposition and sample entropy. The proposed model is further validated for eight wind speed datasets and results reveal the superiority of the model when compared to benchmark models. However, the only limitation of this model is the time consumed in intermediate stages. Three variables,wind speed, electrical load and electricity price are predicted using an Elman neural network (ENN) whose weights and hyperparameters are optimized by a modern dragonfly algorithm as presented by Wang et al. \cite{Wang2018}. Since the time-series for the three variables is highly non-linear, signal processing techniques such as wavelet transform, empirical mode decomposition and ensemble empirical mode decomposition are widely applied in the allied areas of forecasting. 

Du et al. presented a multi-step ahead prediction based on a Whale optimization algorithm (WOA)-LSSVR technique and have applied the same to forecast wind speed, electricity price and electrical load \cite{Du2018}. The proposed technique is validated for six datasets from Singapore, China and Australia. Existing benchmark regressors like Generalized regression neural network (GRNN) and Back propagation neural network (BPNN) are used for a comparative analysis. In terms of performance metrics like root mean squared error and mean absolute error the proposed WOA-LSSVR model outperforms GRNN and BPNN. 

Wang et al. presented a hybrid wavelet neural network (WNN) model based  multi-objective sine-cosine algorithm (MOSCA) optimization technique \cite{Wang2018}. The optimization problem is based on multi-objective sine-cosine functions. The candidate solutions are first initialized with some value and are allowed to converge or diverge in a given search space. A modified complementary ensemble empirical mode decomposition (MCEEMD) is implemented in order to solve the issues posed by simple pre-processing techniques such as ensemble empirical mode decomposition (EEMD). The model proposed by Wang et al. is also evaluated for assessing the robustness and stability for wind speed prediction as a highly non-linear time-series can cause difficulties in accurate prediction. The model based on MOSCA-WNN is applied to predict each sub-series and the results are aggregated. The proposed model is compared with WNN, GRNN, ARIMA and Persistence model. 

\section{Conclusions}
  
In this manuscript, we discuss primarily machine learning and hybrid wind forecasting techniques and its variants for regression analysis. These supervised learning models are useful for wind speed time-series datasets from different places around the globe to assess the performance of each model. In terms of accuracy, machine learning model based on supervised learning give good generalization performance which is the ability of a machine learning model to adapt to an unseen data. However, the need for ensemble based forecasts arrives from the problem of over-fitting that occurs in models like support vector regression, multiple linear regression and neural networks.  Further, the superiority of hybrid forecasting models has led to an increased research in the field of wind speed forecasting.

\bibliographystyle{IEEEtran}
\bibliography{bibfileforpaper}

\end{document}